\documentclass[prl,twocolumn,showpacs,preprintnumbers,amsmath,amssymb,superscriptaddress,floatfix]{revtex4}

\usepackage{graphicx}
\usepackage{dcolumn}
\usepackage{bm}

\newfont{\boldit}{cmbxti10}
\newfont{\boldslant}{cmbxsl10}
\newcommand{\nab}{\mbox{\boldmath $\nabla$}}
\begin{document}

\title{Viscoelastic Behavior of Solid $^4$He}

\author{C.-D. Yoo}
\author{Alan T. Dorsey}

\affiliation{Department of Physics, University of Florida, P.O. Box
118440, Gainesville, FL 32611-8440}

\date{\today}

\begin{abstract}

Over the last five years several experimental groups have reported
anomalies in the temperature dependence of the period and amplitude
of a torsional oscillator containing solid $^4$He. We model these
experiments by assuming that $^4$He is a viscoelastic solid---a
solid with frequency dependent internal friction. We find that while
our model can provide a quantitative account of the dissipation
observed in the torsional oscillator experiments, it only accounts
for about 10\% of the observed period shift, leaving open the
possibility that the remaining period shift is due to the onset of
superfluidity in the sample.

\end{abstract}

\pacs{
67.80.-s, 
67.40.-w, 
46.35.+z} 

\maketitle

In 2004 Kim and Chan \cite{kim04.1,kim04.2} reported anomalies in
the resonant period of a torsional oscillator (TO) containing solid
$^4$He. With exquisite sensitivity they detected a reproducible
decrease in the oscillator period upon lowering the temperature
below 200 mK. Subsequent experiments in several laboratories
\cite{kim05,kim06,clark07,rittner06,kondo06,penzev07,aoki07,rittner07,kim08}
have now reproduced these results. The size of the effect depends on
the sample preparation, pressure, and concentration of $^3$He
impurities. References \cite{prokofev07} and \cite{balibar08} review
the recent theoretical and experimental work.

A natural interpretation of the TO anomalies is the onset of the
elusive and long-anticipated supersolid phase of matter
\cite{chester70,andreev69,leggett70}. In a supersolid, superfluidity
coexists with the crystalline order of a solid; one expects a
supersolid to exhibit superflow, much like a superfluid. Leggett
\cite{leggett70} proposed that the superflow is best detected by
searching for ``nonclassical rotational inertia''; a superfluid
condensate would remain at rest and not participate in rotation, and
the resulting mass decoupling would reduce the rotational inertia
and decrease the resonant period of oscillation. While compelling,
the supersolid interpretation of these experiments has yet to be
corroborated by other measurements, such as the response to pressure
differences \cite{day06}. Moreover, Day and Beamish \cite{day07}
reported a pronounced increase in the shear modulus of $^4$He at
temperatures below 200 mK, with a dependence on measurement
amplitude and $^3$He impurity concentration similar to the TO
anomalies. Their results suggest that changes in the shear modulus
might be intimately related to the TO anomalies.

This Letter presents a detailed discussion of the mechanical
response of a TO containing a viscoelastic solid. We build on
earlier work by Nussinov \emph{et al}.\ \cite{nussinov07}, who
correctly identified a ``back action'' term in the TO equation of
motion that represents the dynamical effect of the solid helium on
the torsion cell. However, in contrast to Nussinov \emph{et al}.\ we
find no need to assume that the solid helium behaves as a glass.
Instead, with a few carefully stated assumptions, we find that we
can model the solid helium as a classical viscoelastic solid---i.e.,
a solid with internal friction. The TO period shift and dissipation
peak are naturally related to the real and imaginary parts of the
frequency-dependent shear modulus of the solid. We use our results
to fit the dissipation peak reported in Clark \emph{et al}.
\cite{clark07}, and extract a temperature dependent time scale
$\tau(T)$ from the data. With all of the phenomenological parameters
determined, we find that we are only able to account for about 10\%
of the period shift reported in Ref.~\cite{clark07}, leaving open
the possibility that the remaining shift is due to the onset of
superfluidity (or supersolidity) in these samples.

Following Nussinov \emph{et al}.\ \cite{nussinov07}, we assume that
the empty torsion cell is perfectly rigid, with a moment of inertia
 $I_\text{osc}$ about its rotation axis. For small angular
displacements $\theta$ the torsion rod provides a restoring torque
$-\alpha \theta$, with $\alpha$ the torsional spring constant. There
is also a damping torque $-\gamma_\text{osc} d\theta/dt$, with
$\gamma_\text{osc}$ a temperature-dependent dissipation coefficient.
The cell is driven by an external driving torque of $\tau_\text{ext}
(t)$. Finally, the solid helium inside the torsion cell exerts a
moment $M(t)$ on the cell. The equation of motion for the cell is
then
\begin{equation}
\bigg( I_\text{osc} \frac{d^2}{d t^2} + \gamma_\text{osc} \frac{d}{d
t} + \alpha \bigg) \theta(t) = \tau_\text{ext}(t) + M(t),
\end{equation}
with $M(t) = \int dt' g(t-t') \theta(t')$ for a linear system that
is invariant under time translations. We can Fourier transform the
equation of motion to find the response function $\chi(\omega) =
\theta(\omega)/\tau_\text{ext}(\omega)$ of the TO:
\begin{equation}
\chi^{-1}(\omega) = -I_\text{osc}\omega^2 -i\gamma_\text{osc}\omega
+ \alpha - g(\omega) . \label{susceptibility}
\end{equation}
The response function is of fundamental importance in interpreting
the TO experiments, as its poles give the resonant frequencies and
their quality factors.

The complex back action term $g(\omega)$ contains \emph{all} of the
information about the dynamical response of the solid helium, and
modeling this quantity is the focus of the remainder of the paper.
Before delving into calculational details we can make a few general
statements about $g(\omega)$. First, causality requires $g(z)$ to be
analytic in the upper half of the complex $z$-plane, so the real and
imaginary parts of $g(z)$ satisfy Kramers-Kronig relations for
$z=\omega$. Second, if the helium behaves as a perfectly rigid,
normal solid, $g(\omega)=I_\text{He}\omega^2$, with $I_\text{He}$
the rigid body moment of inertia of the helium. As we will show
below, a finite shear modulus produces corrections to this result
which vanish with a \emph{higher} power of $\omega$ at low
frequencies. Therefore $g(\omega=0)=0$; the back action only
modifies \emph{dynamical} quantities. At low frequencies the
``glassy'' model for $g(\omega)$ proposed by Nussinov \emph{et al}.\
 \cite{nussinov07} [see their Eq. (20)] goes to a constant and
induces an unphysical shift in the static spring constant $\alpha$.

To calculate the back action term we first need to model the
properties of the helium in the torsion cell. To keep the
description as general (and simple) as possible we model the helium
as an elastic continuum, with an equation of motion
\begin{equation}
\rho \partial^2_t u_i = \partial_j \sigma_{ij},
\label{equation_motion}
\end{equation}
where $\rho$ is the density, $u_i$ is the $i^{\rm th}$ component of
the displacement field and $\sigma_{ij}$ is the stress tensor. The
stress tensor has both a reversible piece $\sigma_{ij}^r$ and a
dissipative piece $\sigma_{ij}^d$, with $\sigma_{ij} = \sigma_{ij}^r
+ \sigma_{ij}^d$ \cite{landau}. For a linear medium the reversible
stress is linearly related to the strain tensor $u_{ik} = \left(
\partial_k u_i + \partial_i u_k \right)/2$ by
$\sigma_{ij}^r  = \lambda_{ijkl} u_{kl}$, where $\lambda_{ijkl}$ is
a fourth-rank tensor of elastic coefficients. For a uniaxial crystal
such as the hcp phase of $^4$He there are 5 independent components
of $\lambda_{ijkl}$; however, to simplify the analysis we assume
that the helium can be modeled as an isotropic elastic solid, with
$\sigma_{ij}^r = 2\mu_0 u_{ij} + \lambda_0 \delta_{ij} u_{kk}$,
where $\lambda_0$ and $\mu_0$ are the Lam\'e coefficients (with
$\mu_0$ the shear modulus). The dissipative part of the stress
tensor, which describes the \emph{internal friction} of the solid,
must be odd under time reversal, and can only depend on gradients of
the velocity $v_i = \partial_t u_i$ \cite{landau}. For a linear
medium $\sigma_{ij}^d = \eta_{ijkl} v_{kl}$, where $\eta_{ijkl}$ is
the viscosity tensor of the solid and $v_{ik} \equiv (\partial_i v_k
+
\partial_k v_i)/2$. Again assuming an isotropic medium,
$\sigma_{ij}^d = 2\eta v_{ij} + (\zeta - 2\eta/3) \delta_{ij}
v_{kk}$, with $\eta$ and $\zeta$ the shear and bulk viscosities of
the solid. With these simplifications the Fourier transformed
equation of motion becomes
\begin{equation}
-\rho \omega^2 {\bf u} = B(\omega) \nab (\nab \cdot {\bf u}) - \mu
(\omega) \nab\times\nab\times {\bf u}, \label{equation_motion_2}
\end{equation}
with $B(\omega)= \lambda_0 + 2\mu_0 - i\omega (\zeta +4 \eta/3)$ and
$\mu(\omega) = \mu_0 - i\omega \eta \equiv \mu_0 ( 1 - i\tau
\omega)$, with timescale $\tau = \eta/\mu_0$. In passing we note
that this model, known as the Kelvin-Voigt model, is among the
simplest of viscoelastic models---a single ``spring'' (the
elasticity) is in parallel with a single ``dashpot'' (the
viscosity). More elaborate models, involving series and parallel
combinations of springs and dashpots, can produce a shear modulus
$\mu(\omega)$ with a more complicated frequency dependence. For an
example of a similar analysis for colloidal crystals, see
Ref.~\cite{phan99}.

The next step is to determine the response of the helium inside the
torsion cell to the rotation of the cell. For simplicity we will
present results for a torsion cell that is an infinitely long
cylinder of radius $R$; the generalizations to an annular geometry
or a cylinder of height $h$ are straightforward \cite{yoo08}. If we
assume the helium is in perfect contact with the walls of the
torsion cell (no-slip boundary conditions), and the torsion cell
oscillates about its azimuthal axis with a frequency $\omega$ and
amplitude $\theta_0$, then the induced displacements in the helium
have the form ${\bf u} = u_\theta(r)e^{-i\omega t}\, \hat{\theta}$.
Substituting this into Eq.~(\ref{equation_motion_2}) and solving the
differential equation with the no-slip boundary condition
$u_{\theta}(r=R)= R \theta_0 $, the solution that is finite at $r=0$
is
\begin{equation}
u_\theta (r) = R \theta_0 \frac{J_1(k r)}{J_1(k R)},
\end{equation}
where $k^2 = \omega^2 \rho / \mu(\omega)$ and $J_1(z)$ is the Bessel
function of order 1. In this geometry the torsion cell only induces
shearing displacements in the helium.

The final step of the calculation is to determine the moment that
the oscillating helium exerts back on the torsion cell. The only
non-vanishing component of the stress tensor is $\sigma_{\theta r}=
\mu(\omega) (\partial_r - 1/r) u_\theta(r)$; evaluating this on the
surface of the cylinder, integrating over the area of the cylinder
to obtain a force, and then multiplying by the radius to obtain a
torque, we find the moment
\begin{equation}
M(t) = - \theta_0 \omega^2 I_\text{He} \frac{4 J_2(kR)}{kR J_1(kR)}
e^{-i \omega t}, \label{torque}
\end{equation}
where  $I_\text{He} = \pi \rho h R^4 / 2$ is the rigid body moment
of inertia for the helium inside the torsion cell. In terms of the
back-action term $g(\omega)$ defined in Eq. (\ref{susceptibility}),
\begin{equation}
g(\omega) =  I_\text{He}\omega^2 + I_\text{He}\omega^2
\tilde{g}(kR), \label{backaction1}
\end{equation}
where the function
\begin{equation}
\tilde{g}(x) = \frac{4 J_2(x)}{x J_1(x)} -1.
\label{backaction2}
\end{equation}
is the correction to the rigid body result due to the finite shear
modulus of the helium.

To simplify our result, we note that for a typical TO the speed of
transverse sound $c_T = \sqrt{\mu_0/\rho}\sim 270\ \text{m/s}$, the
frequency $f\sim 10^3\ \text{s}^{-1}$, and the radius $R\sim 0.5\
\text{cm}$, so that $|k|R =2\pi f R/c_T \sim 0.1$. Therefore we can
safely expand Eq. (\ref{backaction2}) for small $x$, with the result
$\tilde{g}(x) \simeq x^2 / 24$; then Eq.~(\ref{susceptibility})
becomes
\begin{equation}
\chi^{-1}(\omega) \simeq -I_\text{tot}\omega^2
-i\gamma_\text{osc}\omega + \alpha - \frac{\rho R^2 \omega^4
I_\text{He} F(R/h)}{24 \mu (\omega)}, \label{susceptibility.1}
\end{equation}
where $I_\text{tot} = I_\text{He} + I_\text{osc}$. We see that the
last term in Eq.~(\ref{susceptibility.1}) is the correction due to a
finite shear modulus; for a perfectly rigid body, $\mu\rightarrow
\infty$ and this term vanishes. Also, in this last term we have
introduced a function $F(x)$ to describe the effect of a finite
cylinder height $h$ \cite{yoo08}; this function only depends on the
aspect ratio $x=R/h$, with the explicit form
\begin{equation}
F(x) = - \frac{192 }{\pi^4 x^2} \sum_{m=1}^\infty \frac{1}{(2m-1)^4}
\tilde{g} \bigg( i(2m-1)\pi x \bigg).
\end{equation}
For the infinite cylinder $F(0)=1$, and more generally $0 \le F(x)
\le 1$. For large $x$, $F(x) \simeq 2 / x^2 - 744 \zeta(5) / \pi^5
x^3$ with $\zeta(n)$ being Riemann-Zeta function. In the particular
case $h=R$, $F(1)=0.527$.

We now examine the effect of the viscoelasticity of the helium on
the period and $Q$-factor of the oscillator by finding the poles of
the response function, Eq.~(\ref{susceptibility.1}) (our analysis is
similar to the procedure performed in Nussinov~\emph{et al}.\
\cite{nussinov07}). Since $|k| R = \omega_0 R/c_T \sim 0.1$ and
$I_\text{He} / I_\text{tot} \sim 10^{-3}$, we can treat the fourth
term in Eq.~(\ref{susceptibility.1}) as perturbation about the
rigid-body behavior, which has a resonant frequency $\omega_0 =
\sqrt{\alpha / I_\text{tot}}$ and dissipation $Q_0^{-1} =
\gamma_\text{osc}/\sqrt{I_\text{tot} \alpha}$. Expanding the poles
about $\omega_0$ and $Q_0^{-1}$, and recalling that $\mu (\omega) =
\mu_0 ( 1-i\tau \omega)$, we obtain the fractional period shift
$\Delta P/P_0 = (P-P_0)/P_0$ and the shift in the dissipation
$\Delta Q^{-1} = Q^{-1} - Q_0^{-1}$
\begin{equation}
\frac{\Delta P}{P_0} =  A \,\mathrm{Re}
\left[\frac{\mu_0}{\mu(\omega_0)}\right] = A \frac{1}{1 +
(\tau\omega_0)^2}, \label{period}
\end{equation}
\begin{equation}
 \Delta Q^{-1} = 2A \,\mathrm{Im}
\left[\frac{\mu_0}{\mu(\omega_0)}\right] = 2 A \frac{  \tau
\omega_0}{1 + (\tau \omega_0)^2}, \label{dissipation}
\end{equation}
where the dimensionless amplitude $A$ is given by (recall
$c_T=\sqrt{\mu_0/\rho}$)
\begin{equation}
A = \frac{F(R/h)}{48} \frac{ I_\text{He} }{I_\text{tot}}
\left(\frac{\omega_0 R}{c_T}\right)^2 .
\end{equation}
While $A$ depends on the material parameters and the sample
geometry, \emph{it is independent of the relaxation time $\tau$}.
For a typical TO the amplitude $A$ is of order $10^{-6} - 10^{-7}$,
so the resulting shifts are small, as assumed. As we will show
below, amplitudes in this range can quantitatively fit the
dissipation peak, but are a factor of 10 too small to fit the period
shift of the TO results.

The simple Lorentzian form of these results suggests a strategy for
fitting the TO experimental data. Notice that $\Delta Q^{-1}$ will
have a peak as a function of temperature $T$ if the relaxation time
$\tau$ passes through the time scale $\omega_0^{-1}$ as the
temperature is lowered; the peak occurs at $T^*$ such that $\omega_0
\tau(T^*) = 1$, and at this temperature $\Delta Q^{-1} (T^*) = A$.
Therefore, $A$ can be directly determined from the peak value of the
dissipation; we can then solve Eq.~(\ref{dissipation}) for $\omega_0
\tau$ in terms of $\Delta Q^{-1}(T)/A$ as a function of temperature,
allowing us to determine $\tau(T)$. Having determined both $A$ and
$\tau$, we can then calculate the period shift $\Delta P/P_0$ as a
function of temperature, \emph{with no additional fitting
parameters}. In passing we note that the temperature dependence of
the shear modulus itself \cite{day07} has a much smaller effect on
the period shift and dissipation than the temperature dependence of
the relaxation time $\tau$.

\begin{figure}[htbp]
\centering
\includegraphics[width=8.5cm]
{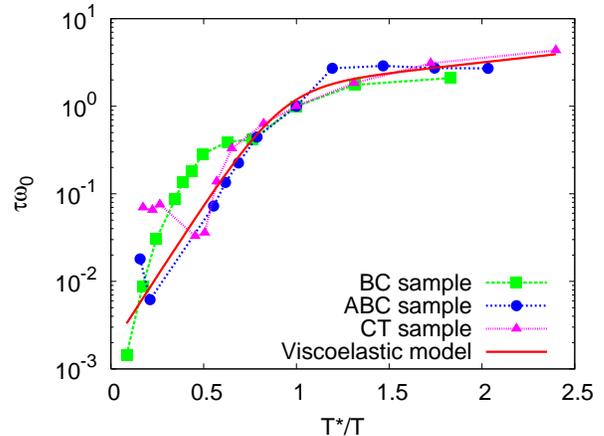} \caption{(color online). Relaxation times
derived from $\Delta Q^{-1}(T)$ of BeCu TO of Clark~{\it et al.\
}~\cite{clark07}. Green squares are for the blocked capillary (BC)
sample, blue circles the annealed blocked capillary (ABC) sample,
and pink triangles the constant temperature (CT) sample. The red
line was found by fitting to all three samples and using $\tau(T)
\omega_0 = a\exp(b T^{*}/T) / [1 + c\exp(d T^{*}/T)]$ with fitting
parameters $a=1.75\times10^{-3}$, $b=7.55$, $c=1.62\times10^{-3}$
and $d=7.01$. } \label{figure1}
\end{figure}
\begin{figure}[htbp]
\centering
\includegraphics[width=8.5cm]
{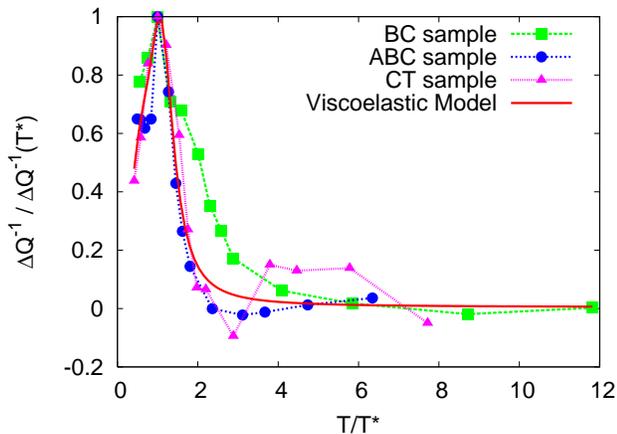} \caption{(color online). Peaks in the inverse
of the quality factor of BeCu TO from Clark~{\it et al.\
}~\cite{clark07}. Green squares are of the blocked capillary (BC)
sample, blue circles the annealed blocked capillary (ABC) sample,
and pink triangles the constant temperature (CT) sample. The red
line is of the viscoelastic model with the change in $Q$-factor
Eq.~(\ref{dissipation}) and the derived relaxation time
Eq.~(\ref{relaxation.time}).} \label{figure2}
\end{figure}
\begin{figure}[htbp]
\centering
\includegraphics[width=8.5cm]
{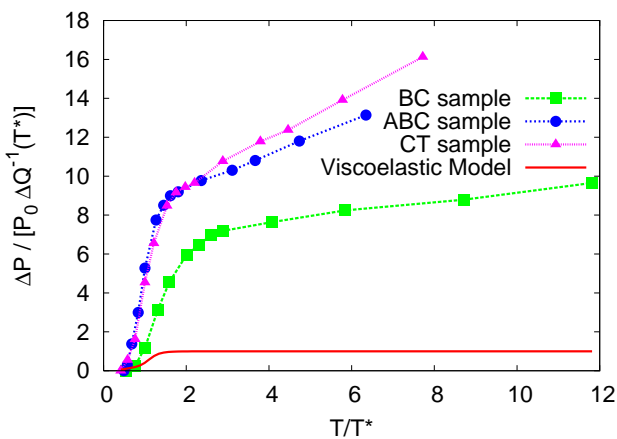} \caption{(color online). Resonant period
shifts of BeCu TO from Clark~{\it  et al.\ }~\cite{clark07}. Green
squares are of the blocked capillary (BC) sample, blue circles the
annealed blocked capillary (ABC) sample, and pink triangles the
constant temperature (CT) sample. The red line is of the
viscoelastic model with the resonant period shift Eq.~(\ref{period})
and the derived relaxation time Eq.~(\ref{relaxation.time}).}
\label{figure3}
\end{figure}

In Fig.~\ref{figure1} we show the relaxation time obtained from the
measured $Q$-factor of the BeCu TO used in Ref.~\cite{clark07} (to
collapse all of the data onto a single curve we have scaled the
temperature by the peak temperature $T^*$). On this log plot we
clearly see activated behavior both below and above $T^*$, but with
different activation energies. To account for this behavior, we will
assume that the relaxation time $\tau$ has the functional form
\begin{equation}
\tau(T) = \frac{\tau_0 \exp(E_0/T)}{1+\delta\exp(E_1/T)}.
\label{relaxation.time}
\end{equation}
For BeCu TO data of Ref.~\cite{clark07} we obtain $\tau_0 = 260$ ns,
$\delta = 1.62\times10^{-3}$, $E_0 = 7.55 T^{*}$, and $E_1 = 7.01
T^{*}$. At high temperatures, the activation energy for the blocked
capillary sample of BeCu TO of Ref.~\cite{clark07} is found to be $E
= 260.4$ mK (417.5 mK for the annealed blocked capillary sample and
341.4 mK for the constant temperature sample), and at low
temperatures $E = 18.6$ mK (29.8 mK and 24.4 mK, respectively). By
using the derived relaxation time $\tau$ the fits to the dissipation
peak and period shift data from Ref.~\cite{clark07} are shown in
Figs.~\ref{figure2} and \ref{figure3}. Having fit to the dissipation
peak, we see that the same set of parameters accounts for only 10\%
to 20\% of the period shift observed in these experiments.


In summary, we investigated the viscoelastic behavior of solid
$^4$He at low temperatures. The response of a viscoelastic solid to
an oscillatory shear stress is determined and used to study the
anomalies in the resonant period and the dissipation of the TO
experiments. Our approach is quite general, invoking the linear
response of the helium together with the simplifying assumption of
isotropy; our framework can be used to study other phenomenological
models for the mechanical behavior of solid helium. For instance, we
could also treat a model with a distribution of relaxation times,
such as a ``glass'' model for the shear modulus \cite{nussinov07} of
the form $\mu(\omega) = \mu_0 (1-i\tau\omega)^\beta$. The simple
single relaxation time Kelvin-Voigt model identifies a time scale
associated with the viscosity of solid $^4$He; upon lowering the
temperature, this relaxation time grows rapidly and eventually
passes through $\omega_0^{-1}$, inducing changes in both the
dissipation and the oscillator period. While the dissipation peak
can be explained completely using the viscoelastic model, the model
accounts for only a fraction of the period shift observed in
Ref.~\cite{clark07} (although fits to some data \cite{rittner06}
yield a smaller discrepancy between the model results and measured
period shifts). As originally suggested \cite{kim04.1,kim04.2}, the
remaining period shift may indeed be due to the onset of some type
of superfluidity in the solid helium.

\begin{acknowledgments}
The authors would like to thank A. Balatsky, J. Beamish, M. Chan, T.
Clark, K. Dasbiswas, P. Goldbart, D. Goswami, M. Graf, D. Huse, H.
Kojima, Y. Lee, M. Meisel, N. Mulders, E. Pratt, J. Toner and J.
West for helpful discussions and comments; and ATD would like to
thank the Aspen Center for Physics, where part of this work was
completed. This work was supported by the National Science
Foundation.
\end{acknowledgments}

\end{document}